\documentclass[useAMS,usenatbib]{mn2e}
\usepackage[dvipsnames]{color}
\usepackage{graphicx}

\def \be{\begin{equation}}
\def \ee{\end{equation}}
\def \bea{\begin{eqnarray}}
\def \eea{\end{eqnarray}}
\def\etal{{et al.\ }}

\title{The role of 3-body H$_2$ formation in the fragmentation of primordial gas}

\author[Jayanta Dutta \etal]
{\parbox{\textwidth}
{Jayanta Dutta$^{1,2}$, 
Biman B. Nath$^1$, Paul C. Clark$^{2,3}$, Ralf S. Klessen$^2$} 
\vspace{0.4cm}\\
\parbox{\textwidth}{
$^1$Raman Research Institute, Sadashiva Nagar, Bangalore 560080, India\\
$^2$Universit\"at Heidelberg, Zentrum f\"ur Astronomie, Institut f\"ur Theoretische  Astrophysik, Albert-Ueberle-Str. 2, 69120 Heidelberg, Germany\\
$^3$School of Physics and Astronomy, Cardiff University, 5 The Parade, Cardiff CF24 3AA, UK\\
}}

\begin{document}

\maketitle

\label{firstpage}

\begin{abstract}
It has been shown that the behaviour of primordial gas collapsing in a dark matter minihalo can depend on the adopted choice of 3-body H$_2$ formation rate. The uncertainties in this rate span two orders of magnitude in the current literature, and so it remains a source of uncertainty in our knowledge of population III star formation. Here we investigate how the amount of fragmentation in primordial gas depends on the adopted 3-body rate. We present the results of calculations that follow the chemical and thermal evolution of primordial gas as it collapses in two dark matter minihalos. 
Our results on the effect of 3-body rate on the evolution until the first protostar forms agree well with
previous studies. However, our modified version of GADGET-2 SPH also includes sink particles, which allows us to follow the initial evolution of the accretion disc that builds up on the centre of each halo, and capture the fragmentation in gas as well as its dependence on the adopted 3-body H$_2$ formation rate. We find that the fragmentation behaviour of the gas is only marginally effected by the choice of 3-body rate co-efficient, and that halo-to-halo differences are of equal importance in affecting the final mass distribution of stars. 

\end{abstract}


\section{Introduction}
\label{sec:introduction}

According to the standard model of the primordial star formation process, the very first 
stars, the so-called Population~III (or Pop.~III)
stars form within dark matter (DM) halos with virial temperature $\sim 1000$ K and masses 
$\sim 10^6$ $M_\odot$ that collapsed at redshift $z \sim$ 25-30 or above
(\citealt{htl96,tegmark97,abn02,g05,bl04,byhm09}). The hydrogen 
atoms in the DM halos combine with the small abundances of free electrons via
$\rm H + e^- \rightarrow H^- + \gamma$, followed by $\rm H^- + H \rightarrow H_2 + e^-$,
where the free electrons act as catalysts and are present as residue from the
epoch of recombination at $z \sim$ 1100 (\citealt{bl01,cf05}).

The typical fractional abundances of H$_2$  
of $\sim 10 ^{-3}$ are sufficient to permit the primordial gas in the minihalos to cool as it collapses (see, e.g.\ \citealt{suny98,yahs03}) 
via H$_2$ rotational and vibrational 
line emission. This occurs until the gas reaches a temperature $\sim$ 200K \citep{on98} at a density of $\sim 10^4$cm$^{-3}$, at which point the H$_2$ energy levels move into the local thermodynamical equilibrium (LTE) with the kinetic temperature of the gas, and the resulting cooling time becomes longer than the free-fall of the gas. The gas now heats up slightly as the collapse proceeds. This transition from cooling to heating with increasing density sets a characteristic Jeans length, allowing the gas to fragment with Jeans mass of $M_{\rm J}$  $\sim$ 1000$M_\odot$ at typical temperatures of $T \sim 200\,$K and number densities $n \sim 10^4\,$cm$^{-3}$
(\citealt{aanz98,abn02,bcl99,bcl02}).

Once the gas reaches a number density of $\sim 10^8$cm$^{-3}$, further H$_2$ formation is possible via the  3-body reactions \citep{pss83}: 
\begin{equation}
\rm H + H + H \rightarrow H_2 + H
\end{equation}
\begin{equation}
\rm H + H + H_2 \rightarrow H_2 + H_2.
\end{equation}
Unfortunately, the rate coefficients for the above set of reactions are very uncertain in the temperature regime which is applicable to Pop. III star formation, with the current estimates in the literature spanning two orders of magnitude (see \citealt{ga08, gs09} for discussion). However, independent of the choice of rate coefficient, calculations of the gravitational collapse of primordial gas show that by a number density of $\sim 10^8$cm$^{-3}$, almost all the atomic hydrogen is converted to H$_2$ (\citealt{yoha06,tao09}). 
 
The sudden formation of $\rm H_2$ has two interesting effects. First, the process is accompanied by heating, since 4.4 eV is released for every H$_2$ molecule that is formed. If the formation rate is fast enough, this heating can momentarily stall the collapse against collapse. On the other hand, the ability of the gas to cool is proportional to (among other things) the fractional abundance of H$_2$, so the more H$_2$ is formed, the more the gas can cool. In practise, the interplay between these two phenomena is complex, and depends crucially on the adopted rates for the 3-body reactions \citep{tcggakb11}. Under certain conditions, a chemo-thermal instability can also occur that may promote the gas to fragment \citep{yoh07,gsb13}. 

In addition to any fragmentation induced by the chemo-thermal instability, recent simulations conclude that the disk around the first primordial protostar becomes gravitationally unstable and forms a multiple system with low-mass protostars instead of a single protostar \citep{cgk08,cgsgkb11b,sgb10,gswgcskb11,sgcgk11,dgck13}. This can occur on timescales as short as 10 to 100 years after the formation of the first object. \citet{cgsgkb11b} show that the fragmentation of this disc results from the fact that the disc is unable to accrete onto the protostar faster than it accretes from the in-falling envelope. 

The study by \citet{tcggakb11} used both SPH and AMR simulations to conclude that the 
choice of 3-body-rate coefficients can introduce significant uncertainty into the 
radial velocity, temperature and accretion rate, and therefore can affect the rate at 
which the protostellar disc is fed. {
This  has recently been verified by 
\citet{bsg14}, who used both old rates and newly calculated improved rate given by 
\citet{forrey13}, to reach similar conclusion as \citet{tcggakb11}. Although they 
found  compressed spiral arms, with strong density contrast between the 
arms and embedding medium, these simulations were unable to follow the evolution 
beyond the formation of the protostar, and so could not address the long-term behaviour 
of the disc structure.} 

However the degree to which the disc fragmentation depends on the chemical 
uncertainties in the 3-body H$_2$ formation rate has never been systematically tested. 
{
Previous studies did not include sink particles, and could not study 
the fragmentation that occurs once the first object is formed.} In this 
work we focus on the effect that the choice of 3-body reaction rate has on the 
fragmentation of the primordial gas. %
\section{Numerical Methodology}

\subsection{Simulation setup}
We investigate two minihaloshalos obtained from the cosmological simulations of \citet{gswgcskb11,gbcgskys12}, which used the hydrodynamic moving mesh code {\em Arepo} \citep{springel10}. We use the snapshots from these simulations at a point when the central number density is just below $10^6$ cm$^{-3}$ -- comfortably before the onset of the 3-body reactions -- as the starting point for this project. Interpreting the mesh-generating points of {\em Arepo} as Lagrangian fluids particles, we then use the {\em Arepo} output as the initial conditions for our GADGET-2 \citep{springel05} implementation. The modifications to the standard GADGET-2 include a time-dependent chemical network for primordial gas, and a sink particles to capture the formation of collapsing protostellar cores \citep{bbp95, jap05}. A fuller description of the code and its features can be found in \citet{cgkb11a, cgsgkb11b}. The conversion from the moving mesh to smoothed-particle hydrodynamics (SPH) formalism is the same as that performed by \citet{sgcgk11}, and indeed the minihalos are identical, both in terms of the baryonic and dark matter content.

The original {\em Arepo} simulations resolve the Jeans length with 128 cells \citep{gswgcskb11}, 
at which point further refinement is deactivated, 
resulting in roughly constant mass particles within central $\approx$ 1000 AU.
However, to investigate the effect of 3-body H$_2$ reactions, here we  
use the intermediate {\em Arepo} snapshots (when the central number density 
$n \leq 10^6$ cm$^{-3}$) as the starting point.
Although the initial size of the minihalos are $\approx$ 3 pc, all the fragmentation and accretion take place in the central region of the halos where the
SPH particle masses are roughly 10$^{-4}$ $M_\odot$. In GADGET-2 simulation the mass resolution for 100 SPH particles (e.g. \citealt{bb97}) is $\approx$ 10$^{-2}$ $M_\odot$.

The chemical and thermodynamical treatment of the gas in the {\em Arepo} simulations of \citet{gswgcskb11} is also identical to those in our GADGET-2 implementation for the range of densities that are studied in this paper. More details of the initial conditions for this study are given below.

Of the various rates available in the literature for the 3-body H$_2$ formation reactions, we choose the extreme rate coefficients: the slowest of which is \citet{abn02} (hereafter ABN02) alongside that of \citet{fh07} which provides the fastest of the rates (hereafter FH07). 
We have used the collisional-dissociation rate in the LTE limit as described in \citet{tcggakb11}.
For our SPH calculation we used a temperature-dependent value calculated via the principle of
detailed balanced. For the ABN02 runs, data from \citet{orel87} are used for low density 
(and low temperature, $T<300$ K) cases, and a power law (described in Table 1) for high density
and temperatures. For FH07 runs, we used the rates given in \citet{fh07}. For the radiative
cooling rate, 
we use Sobolev approximation to calculate it in the optically
thick region, as described in detail in \citet{cgkb11a} and \citet{yoha06}. 
Full details of these rates are given in Table~1.  

\begin{table}
\label{tab:3bh2rates}
\centering
  \begin{tabular}{|ccc|}
    \hline
\hline
 Ref & 3-body H$_2$ formation rates, $k_{\rm 3bh2}$ & Temp range \\
 & (cm$^6$ s$^{-1}$) &  \\ \hline
    \hline
    ABN02 &  1.3 $\times$ 10$^{-32} (T/300)^{-0.38}$  & ($T < 300 K$)\\ 
    ABN02 &  1.3 $\times$ 10$^{-32} (T/300)^{-1.00}$ & ($T > 300 K$)\\ 
    FH07 & 1.44 $\times$ 10$^{-26}/T^{1.54}$ &  \\
    \hline
&H$_2$ collisional dissociation rates 
(cm$^3$ s$^{-1}$) & \\ \hline
    \hline
    ABN02 &  $\frac{1.0670825 \times 10^{-10} \times (T/11605)^{2.012}}{\exp(4.463/T/11605) \times (1 + 0.2472~T/11605)^{3.512}}$ & \\ \\ 
    FH07 &  1.38 $\times$ 10$^{-4} \times T^{-1.025} \exp(-52000/T)$ & \\
    \hline    
  \end{tabular}
\caption{Summary of 3-body H$_2$ formation and dissociation rates adopted here.}
\end{table}

\subsection {Initial Conditions}
The simulations with halo1 start with a maximum central cloud number density ($n$)
10$^{6}$cm$^{-3}$ and minimum $n$ of 71cm$^{-3}$. The initial temperature
of the atomic hydrogen is 470K (max) and 60K (min). Halo1 contains 1030$M_{\odot}$
of gas and 690855 SPH particles. The maximum and minimum mass of SPH particles are
0.16 $M_{\odot}$ and 1.3 $\times 10^{-4}$ $M_{\odot}$.  
This means that the numerical resolution of our simulations for halo1 for 100 
SPH particles is 1.3 $\times 10^{-2}$ $M_{\odot}$.  
The simulations with halo2 starts with a maximum $n$ of 
10$^{6}$cm$^{-3}$ and minimum $n$ of 85cm$^{-3}$. The initial temperature
of the atomic hydrogen is 436K (max) and 54K (min). Halo2 contains 1093$M_{\odot}$
of gas and 628773 SPH particles. The maximum and minimum mass of SPH particles are
0.05$M_{\odot}$ and 1.4 $\times 10^{-4}$ $M_{\odot}$.  
This means that the numerical resolution of our simulations for halo2 for 100 
SPH particles is 1.4 $\times 10^{-2}$ $M_{\odot}$. 
With altogether 6 million particles in the simulation, we resolve the halo gas and its velocity structure significantly better than in the SPH models of our previous study \citep{tcggakb11}. However, we note that none of the existing numerical calculations are able to fully resolve the turbulent cascade in the halo gas and do justice to the  large Reynolds numbers (of $10^9$ or above) that characterise the accretion flow \cite[see, e.g.\ chapter 4 in][]{KG15}.

Each of the minihalos are simulated with the two contrasting 3-body reaction rates, resulting in 4 simulations in total. The different minihalos allow us to distinguish between those features of the high-density collapse/evolution that are caused by the chemo-thermal differences arising from the reaction rates, and those features which result from cosmic variance (i.e. differences caused by the formation of the minihalos).

As collapse progresses in the central region of the cloud, sink particles are 
created once the number density of the gas reaches 5 $\times 10^{13}$cm$^{-3}$, at which 
point the gas has a temperature of around 1000K. The corresponding Jeans mass at 
this density and temperature is 0.06$M_{\odot}$, so our calculations  for both of 
the halos are well resolved. We replace a candidate particle by a sink particle
(see, e.g. \citealt{bbp95,kmk04,jap05}) that can accrete gas particles within 
its accretion radius $r_{\rm acc}$ that we fix at 6 AU, which is the Jeans radius at 
the density threshold for sink creation.   
We also prevent sink particles from forming within $2\, r_{\rm acc}$ of 
one another in order to avoid spurious formation of new sink particles out of 
gas that, in reality, would by accreted by a neighbouring sink particle. 
Lastly, gravitational interactions between sinks, and between the sinks and the 
gas, are softened using a fixed softening parameter of 1.2 AU for the sinks. 
We run our simulations until a mass of $21$ M$_\odot$ has been accreted by
all the sink particles.
The final time scales range between $2500\hbox{--}4000$ yr due to the physical differences between 
the minihalos and the 3-body H$_2$ formation rates.

\label{sec:instability}
\begin{figure}
\centerline{
\includegraphics[width=3.5in]{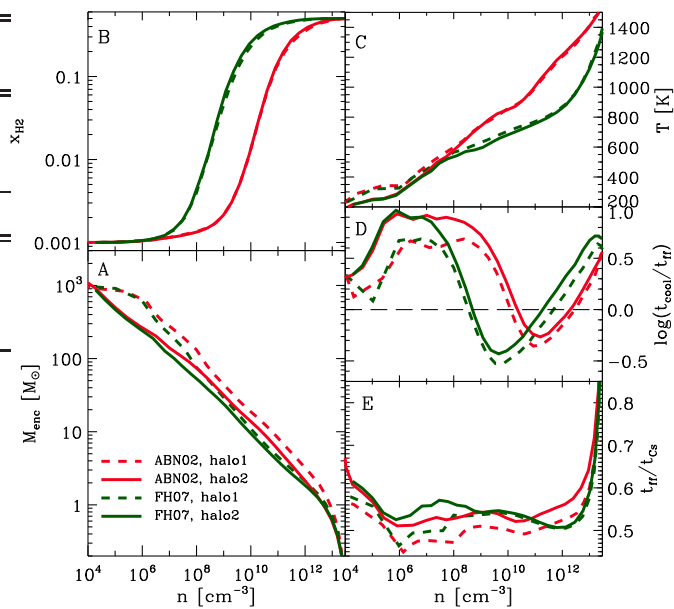}
}
\caption{\label{fig:gas_props} 
Radially binned, mass-weighted averages of physical quantities for different 
3-body H$_2$ formation rate (red: ABN02, green: FH07) of two different cosmological
halos (dotted: halo1, solid: halo2), are compared just before the first protostar
forms. The enclosed mass (A), H$_2$ fraction (B), temperature (C), ratio of 
cooling time to free-fall time (D) and ratio of free-fall time to sound-crossing 
time (E) are plotted as a function of number density. 
}
\end{figure}

\section{Physical conditions just before sink particle creation}

In this section we examine the physical conditions in the gas once the central region has collapsed to a density of $\sim 5 \times 10^{13}$ cm$^{-3}$. The radial profiles of the gas  for the two different halos and two different 3-body rates are shown in Figure \ref{fig:gas_props}. The panels show mass-weighted averages of the properties of individual SPH particles within radial logarithmic bins.

Panel A shows the mass enclosed as a function of the number density, and we see that it displays a rough power-law over most of the density range covered here. This simply follows from the radial power-law in the density which is typical for primordial gas: $n(r) \propto r^{-2.2}$ (e.g. \citealt{abn02,yoha06,yoh08}). We see that there are slight differences between the profiles for the two rates, but these are also comparable to the differences between the two halos.

In panel B, we see the rapid rise in the H$_2$ fraction with increasing density that arises from the 3-body reactions. Here we see a clear difference between the two rates, with the runs with FH07 producing most of the H$_2$ by a density of $10^9$ cm$^{-3}$, while the ABN02 runs are only starting to produce H$_2$ by the time this density is reached. As the rates are a strong function of density, we see very little difference in the H$_2$ fractions between the two halos.

The mean temperature as a function of the number density is shown in panel C. We find that the gas is significantly cooler, up to a factor of 2, around density $\sim 10^{12}$ cm$^{-3}$, in the FH07 halos than in those using ABN02. The differences in temperature-density evolution start around a number density of $10^7$ cm$^{-3}$, once the H$_2$ abundance in the FH07 case has increased by an order of magnitude. Again, we find that there is little difference between evolution of halos 1 and 2 in this panel, implying that the chemical state of the gas is more important for the temperature evolution than any differences in the halo properties. Curves in panels A, B and C are consistent with the results of \citet{tcggakb11}.

\section{Velocity structure and accretion}
\label{sec:accretion}

\begin{figure*}
\centerline
  {
	\includegraphics[width=3.4in]{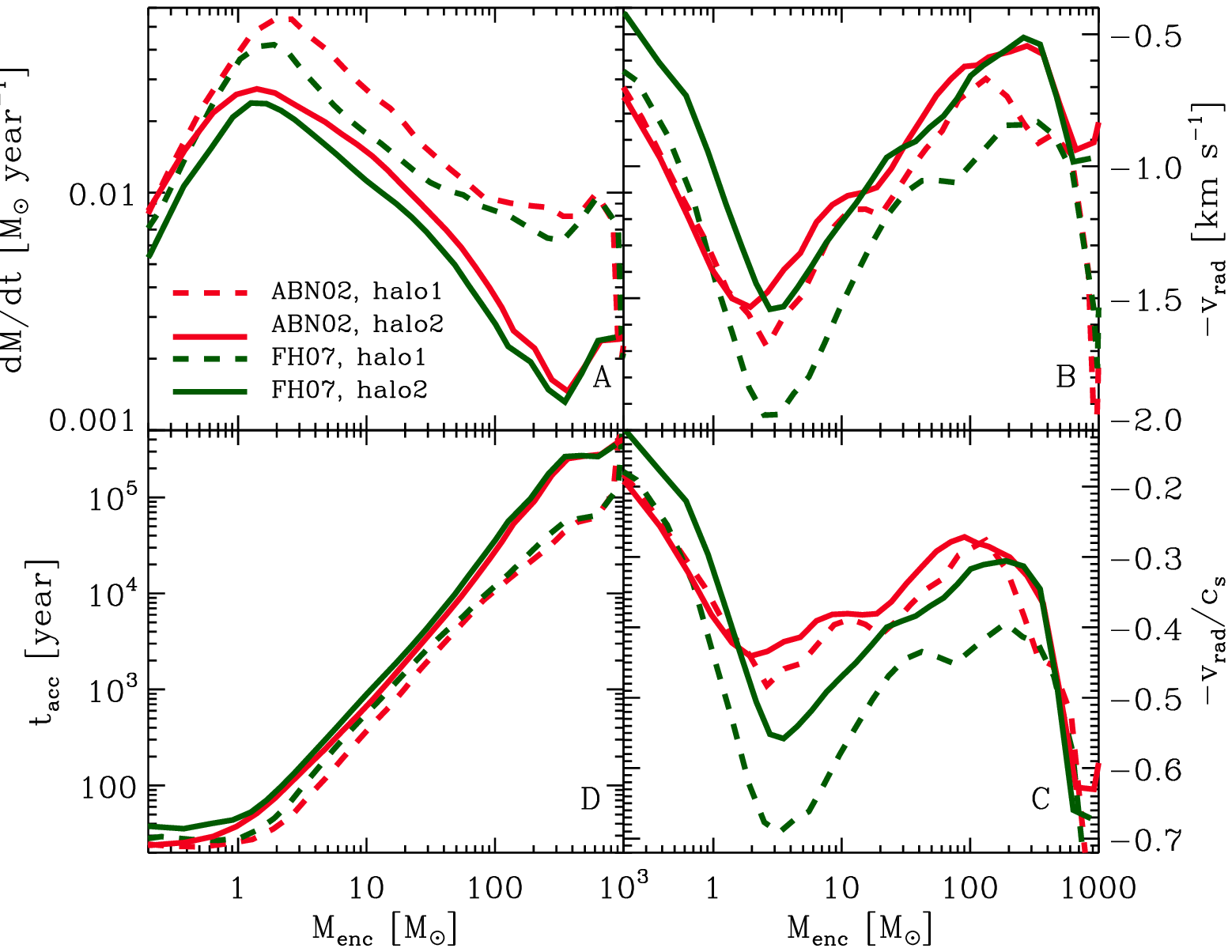}
    \includegraphics[width=3.4in]{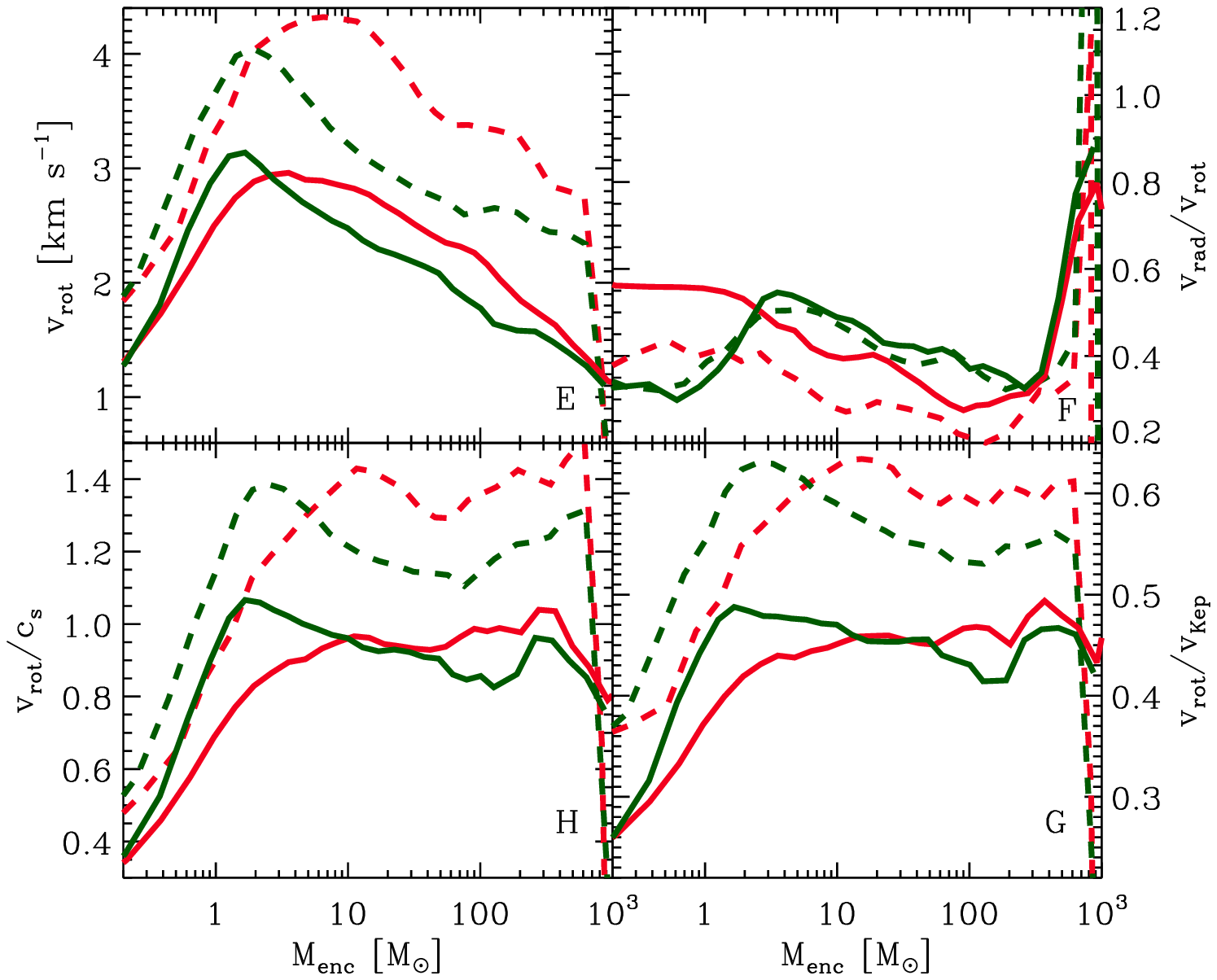}
  }
\caption{\label{vel} Radial logarithmic binned, mass-weighted averages of mass accretion 
rate (A), radial velocities (B),  ratio of radial velocity to sound speed (C), accretion time (D),
rotational velocity (E), ratio of radial to rotational velocity (F), ratio of rotational to Keplerian 
velocity (G), and ratio rotational velocity to sound speed (H)
are plotted as a function of enclosed mass.
The halo to halo variation as well as the uncertainty in the 3-body rate  
introduce an uncertainty in the mass accretion rate, radial and rotational velocity.  
} 
\end{figure*}


In this section we study the difference in mass accretion associated with the
cloud collapse, that can arise due to the choice of 3-body reaction rates. In what follows, we will first study the accretion rate as predicted by the properties of the gas in the clouds just before the first sink particle forms. We will then examine the accretion rate as measured by the sink particles themselves.

The accretion rate can be estimated from the radial gas profiles using,
\begin{equation}
\dot{M}(r) = 4 \, \pi \, r^2 \, \rho(r) \, v_{\rm rad}(r).
\end{equation}
This has been shown to be a good estimate of the accretion rate \citep{cgkb11a}. In addition, one can define an accretion time from,
\begin{equation}
t_{\rm acc} = \frac{M_{\rm enc}(r)}{4 \, \pi \, \rho \, v_{\rm rad} \, r^2}.
\end{equation}

Since accretion is related to infall velocity, we also discuss the corresponding 
radial and rotational velocities. Figure~\ref{vel} shows these physical quantities 
as the function of enclosed mass for all the simulations, using the radial profiles 
in the gas just before the formation of the first sink particle. 
The left panel of Figure~\ref{vel} in the clockwise order, shows the accretion rate,
radial velocity, radial velocity over sound speed and accretion time respectively.
Similarly the right panel of Figure~\ref{vel} shows the rotational velocity, radial 
velocity over rotational velocity, rotational velocity over Keplerian velocity and
rotational velocity over sound speed respectively in the clockwise order.  

We note that the mass accretion rate ($\dot{M}$) for all the simulations has a maximum at an enclosed mass $\sim$ 1\hbox{--}2 $M{_\odot}$. Given that H$_2$ line-cooling becomes optically thick at the corresponding densities for this enclosed mass, it makes sense that $\dot{M}$ for all the simulations converge at this mass scale, as the gas loses its ability to cool efficiently. In general the differences in accretion rates caused by the choice of 3-body formation rate are small, around 25 percent. The differences between the rates in different halos are much larger, at around a factor of 2 for masses below a few tens of solar mass. However we see from the top left panel of Figure \ref{vel} that once the mass approaches 100$\,M{_\odot}$, the difference in the accretion rates can be large, approaching an order of magnitude. We conclude that halo-to-halo variations will play a larger role in the accretion rates than chemical uncertainties.

The choice of 3-body rate coefficient introduces difference in radial velocity of 
$\sim  0.4\hbox{--}0.5$ km s$^{-1}$. 
We also note that the radial velocities for both rates are substantially lower than 
the rotational velocities. The gas particles spiral inwards with trajectories following 
a combination of gravitational attraction and angular momentum conservation. 
All clouds are sub-Keplerian. However, there is a slight tendency for the gas to 
be more rotationally supported in the FH07 runs -- at least for the inner 
10$\,M{_\odot}$ of gas.

We can compare these results with both the AMR and SPH simulations of \citet{tcggakb11}. Our high-resolution study also concludes that the difference in the 3-body $H_2$ formation rates 
significantly affect the dynamical behaviour of the collapsing gas. The uncertainty in the velocity structure has also been investigated by \citet{bsg14} using hydrodynamics code ENZO.
They also found  that there are relatively large differences in the radial velocities of about $2\hbox{--}3$ km s$^{-1}$ depending on the halo distribution as well as 3-body H$_2$ formation rates.

Although there is scatter in the velocities for ABN02 and FH07 rates, the scatter 
also exists in the velocities for both halo1 and halo2. We again conclude that 
the halo-to-halo differences in the velocity structure is, in general, more 
pronounced than that caused by the differences in the chemical reaction rates.

\section{Fragmentation}
\label{sec:fragmentation}

\begin{figure}
\centerline{
\includegraphics[width=3.2in]{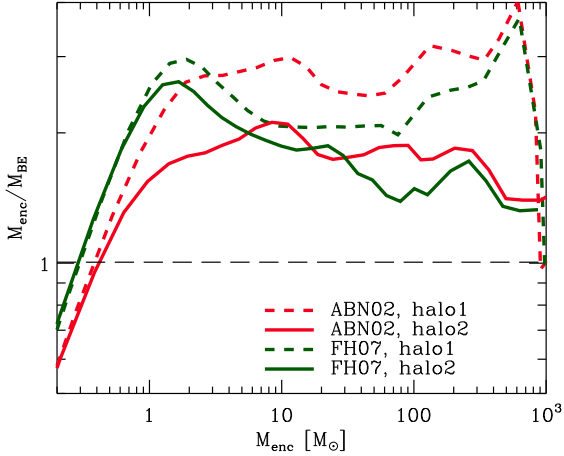}
           }
\caption{\label{MBE} The number of Bonnor-Ebert masses ($M_{\rm enc}/M_{\rm {BE}}$) 
in the central dense volume
is plotted as a function of enclosed mass, just before 
formation of first protostar. The degree of gravitational instability for FH07 within 
the central dense regime is larger than ABN02, indicating larger opportunity for the 
gas to fragment in the FH07 case, than in the ABN02 simulations.}
\end{figure}


We have shown in Section \ref{sec:instability} that the temperature of the gas at high densities, such as those in disc, depends on the choice of adopted 3-body H$_2$ formation rate. In this section, we study how the gas fragments as it evolves beyond the formation of the first collapsing core. Our discussion will first focus on the properties of the gas, to see whether there is any hint of the future evolution already present before the first sink particle forms. We will then compare the properties of the sink particles in the simulations and discuss the differences.

One useful measure of the ability of the gas to fragment is the ratio of the thermal to gravitational energy. This can be represented by the number of Bonnor-Ebert masses ($M_{\rm {BE}}$) contained in the central dense volume (\citealt{ebert1955,bonnor1956}). We therefore investigate the changes in the number (i.e, $M_{\rm {enc}}/M_{\rm {BE}}$) for each of our simulations and how it varies with the enclosed gas mass, similar to the analysis performed by \citet{abn02}. We computed the Bonnor-Ebert mass as the mass-weighted average within the logarithmic radial bin, 
\begin{equation}
M_{\rm{BE}} = 1.18 (c_s^4/G^{3/2})P_{\rm{ext}}^{-1/2} \approx 20 M_\odot T^{3/2} n^{-1/2} \mu^{-2} \gamma^2,
\end{equation}
where $c_s$ is the sound speed, $P_{\rm{ext}}$ is the external pressure that we assume to be equal to the local gas pressure, $\mu$ is the mean molecular weight and $\gamma = 5/3$ is the adiabatic index respectively. 

The results are shown in Fig.~\ref{MBE}.
Note that the ratio $M_{\rm {enc}}/M_{\rm {BE}}$ is equivalent to the ratio of 
fragmentation timescale to that of accretion, $t_{\rm frag}/t_{\rm acc}$.
Once again, these snapshots were taken
immediately  before the formation of the first protostar. The dashed lines
represent the case when fragmentation timescale equals  the accretion timescale.

While we see that there are differences between the mass profiles of the FH07 and ABN02 runs, the effects are, once again, not substantial. However, in the case of $M_{\rm {enc}}/M_{\rm {BE}}$ we see that the FH07 runs have a differently shaped profile from the ABN02 runs (top panel of Fig.~\ref{MBE}). In the FH07 case, the profiles are peaked at characteristic mass of 1-2$\,M_\odot$. In contrast, the ABN02 runs do not display such a peak, and it is less clear whether these runs would favour a particular mass for fragmentation.

If the ratio 
 $M_{\rm {enc}}/M_{\rm {BE}}<1$
the gas enclosed in the shells are accreted faster than it can fragment. As a result fewer new protostars are formed and the available mass contributes to the mass growth of the existing ones \citep{dgck13}. 
In our case we find 
$M_{\rm {enc}}/M_{\rm {BE}}>1$, i.e., the
gas in the shells can fragment faster than it is accreted by the central
dense clump, favouring low-mass protostars. 


\begin{figure*}
\centerline{
\includegraphics[width=3.3in]{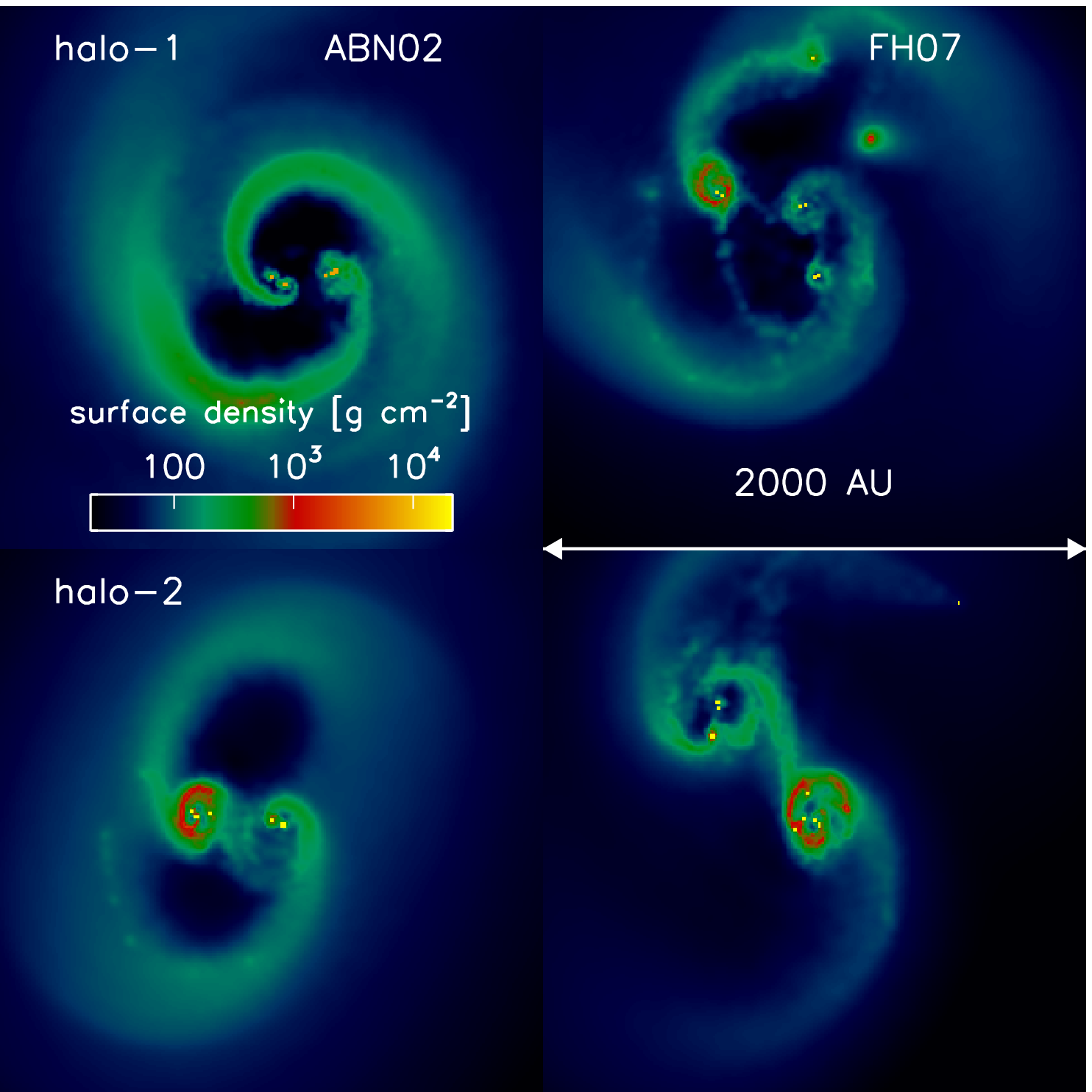}
\includegraphics[width=3.3in]{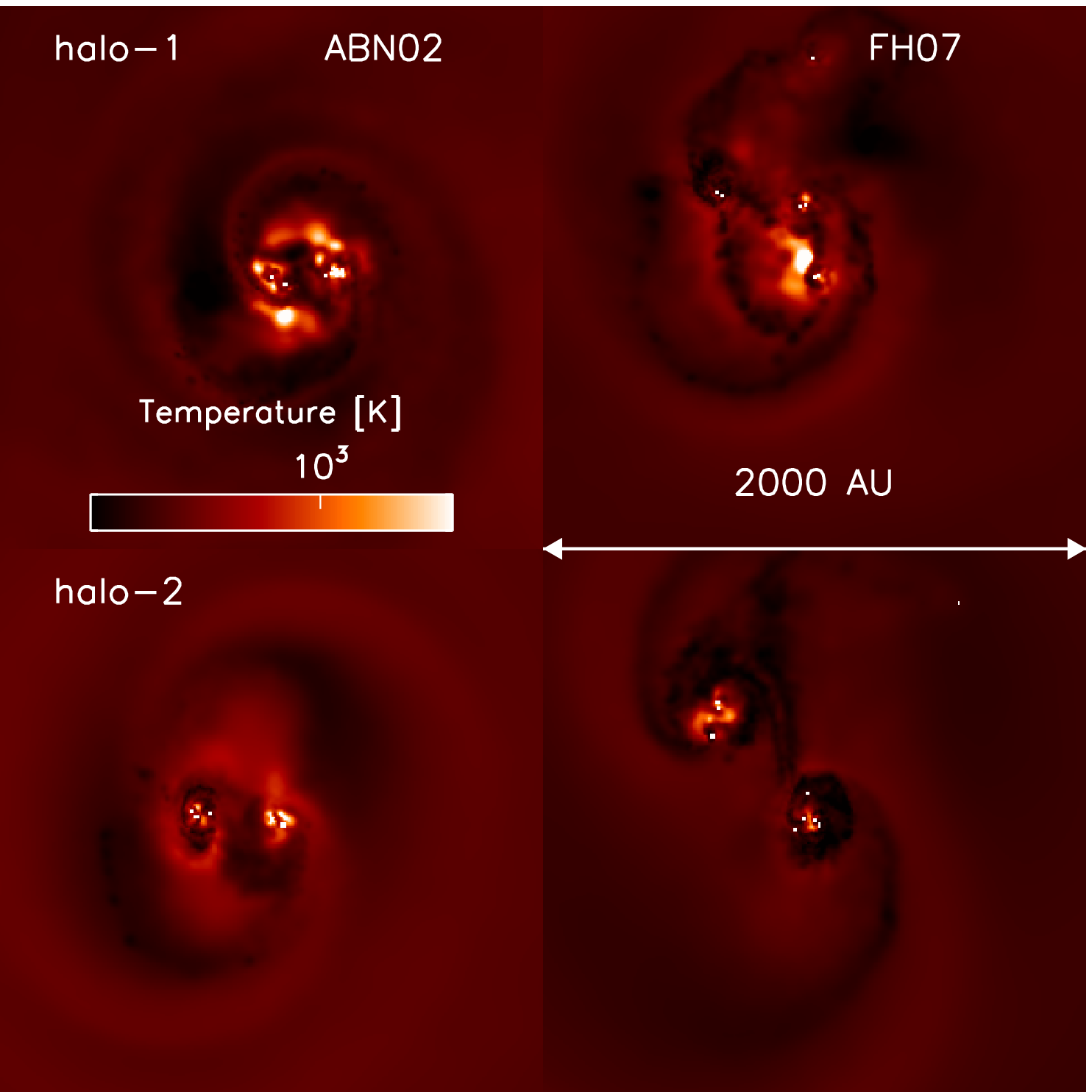}
}
\caption{\label{disk_image} Surface density and temperature in a region of 2000 AU 
centred around first protostar for different 3-body H$_2$ formation rates, are shown when 
the total mass of the sinks reaches $\sim 21\,M_{\odot}$. 
}
\end{figure*}

Therefore by considering the radial density profiles before the onset of sink formation we conclude that gas in FH07 case is perhaps more likely to fragment than in the ABN02 case.
Next we study how the gas fragments at later epochs after formation of the first sink 
particle.
We follow the simulations until $21\,\rm M_{\odot}$ have been converted into, or accreted onto, sink particles. 

We have calculated the total mass accretion rate by all the sinks particles. Although fragmentation may depend on the way sink particles are modelled, our implementation (Section 2.2) takes considerable care to avoid what is often called artificial fragmentation and the spurious formation of sink particles (further details are found in  \citealt{cgsgkb11b}, see also \citealt{BBP95}, \citealt{jap05}, and \citealt{FBCK10}).

The column density and column-weighted temperature distribution in the inner 2000 AU at the end of the simulations are shown in Figure \ref{disk_image}. In all cases we see that the simulations exhibit disc structure on several scales within this central region. It is not
surprising, given the high levels of rotational support seen in right-hand panel in Figure \ref{vel}. We also see that the temperatures in the disc are slightly less in the case with the FH07 rate, due to the increased H$_2$ fraction. However from these images it is clear that all simulations fragment to form a small-$N$ system within the time that $21\,M_\odot$ of material is accreted onto sink particles. 

\begin{figure*}

\setlength{\unitlength}{1.0cm}
\begin{picture}(16, 10) 
\put(-0.5,5.0){\includegraphics[width=5.7cm]{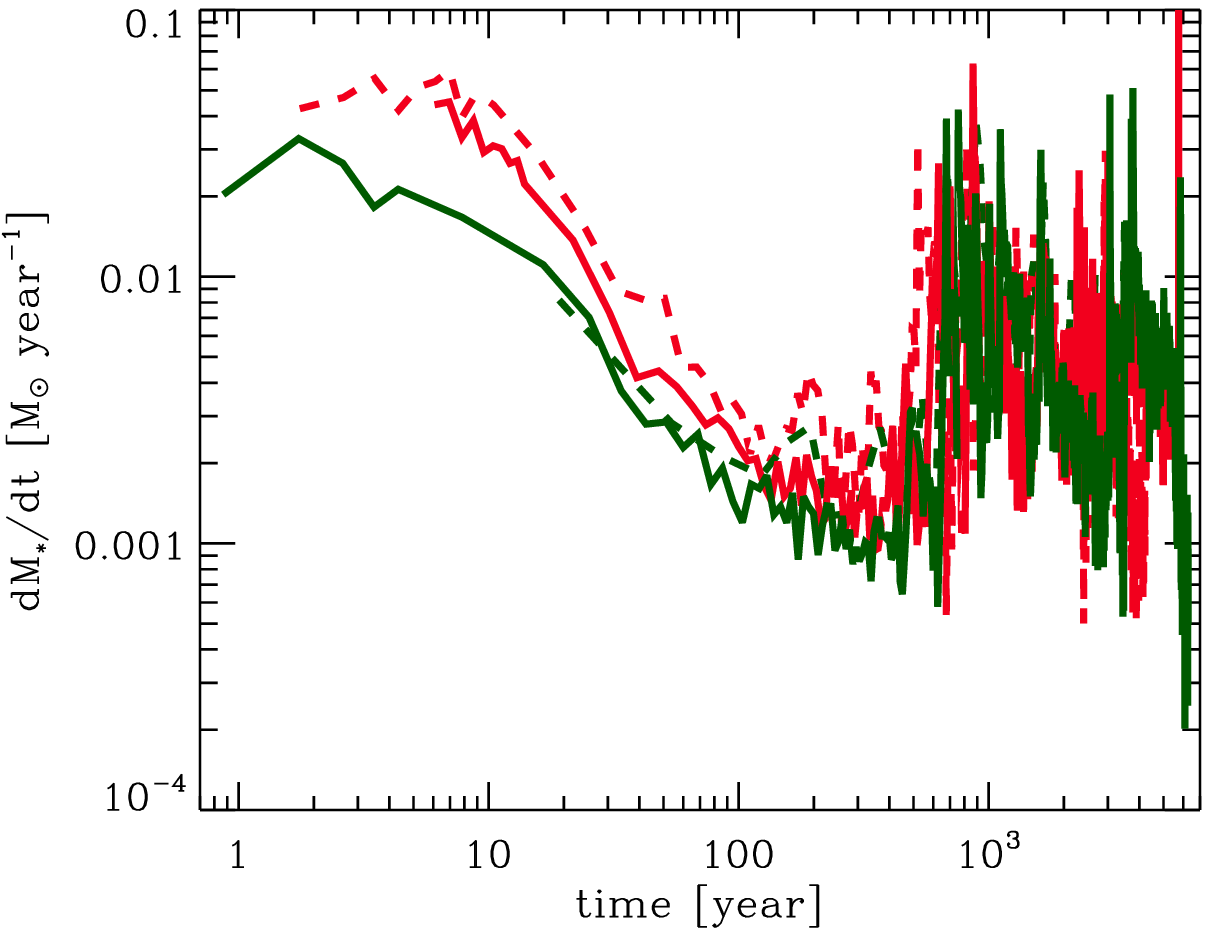}}
\put(5.5,5.0){\includegraphics[width=5.7cm]{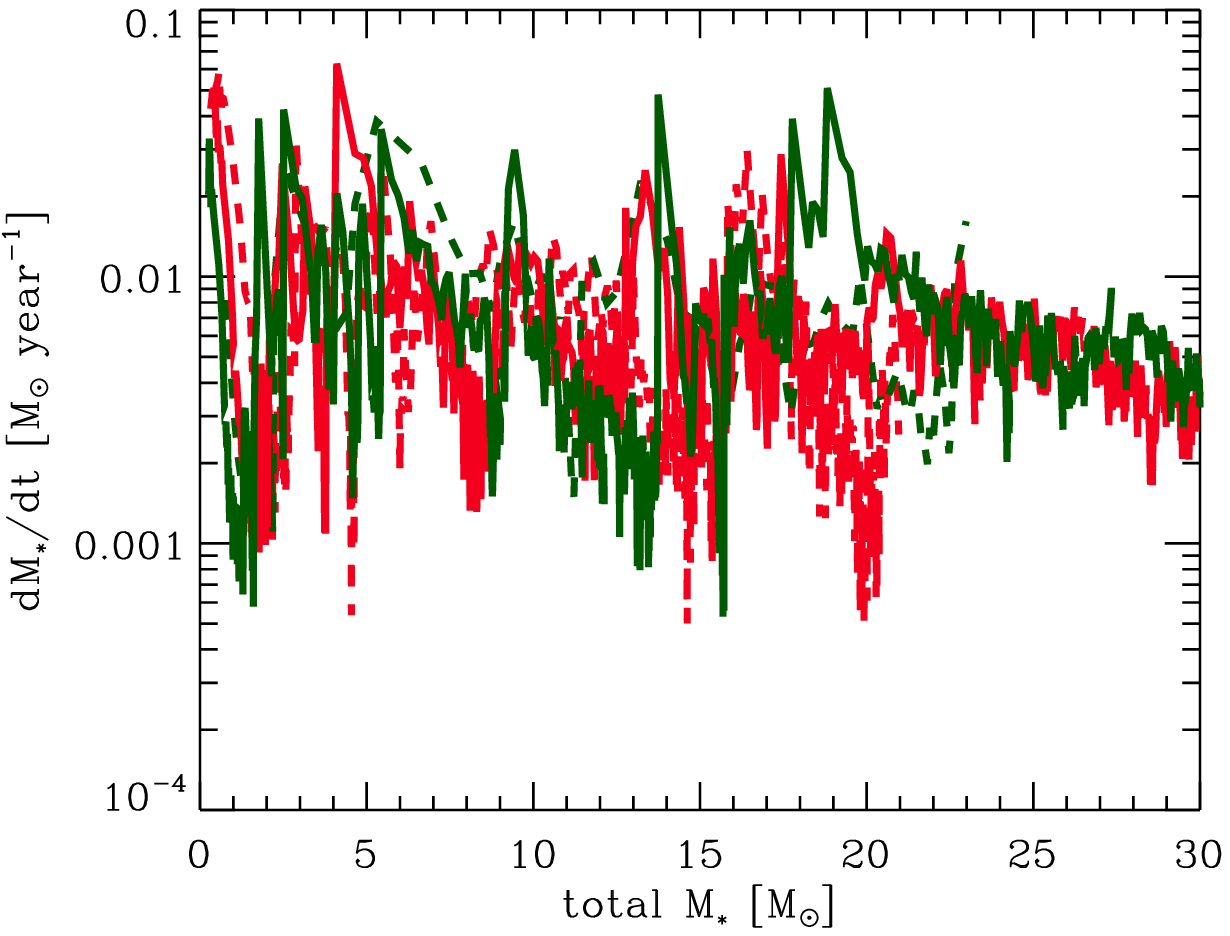}}
\put(11.5,0.0){\includegraphics[width=5.7cm]{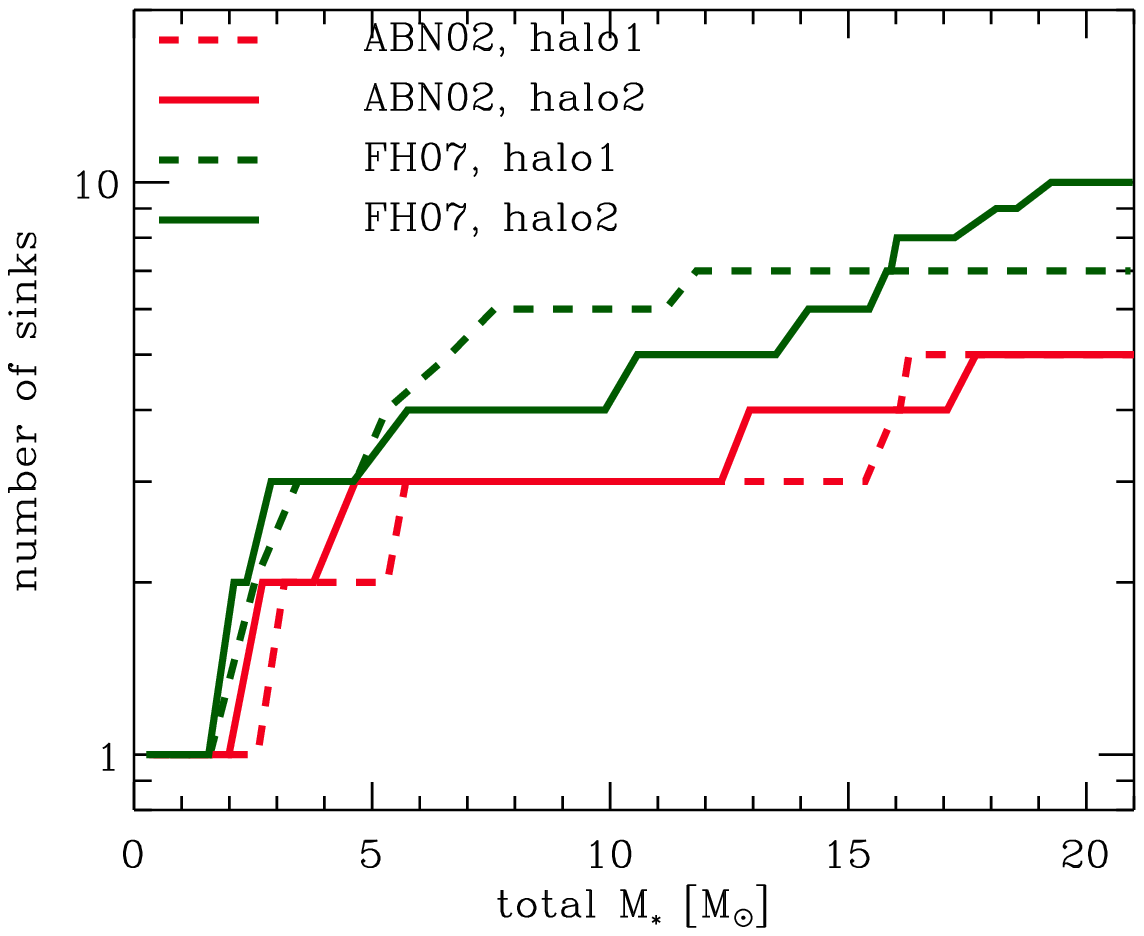}}
\put(-0.5,0.0){\includegraphics[width=5.7cm]{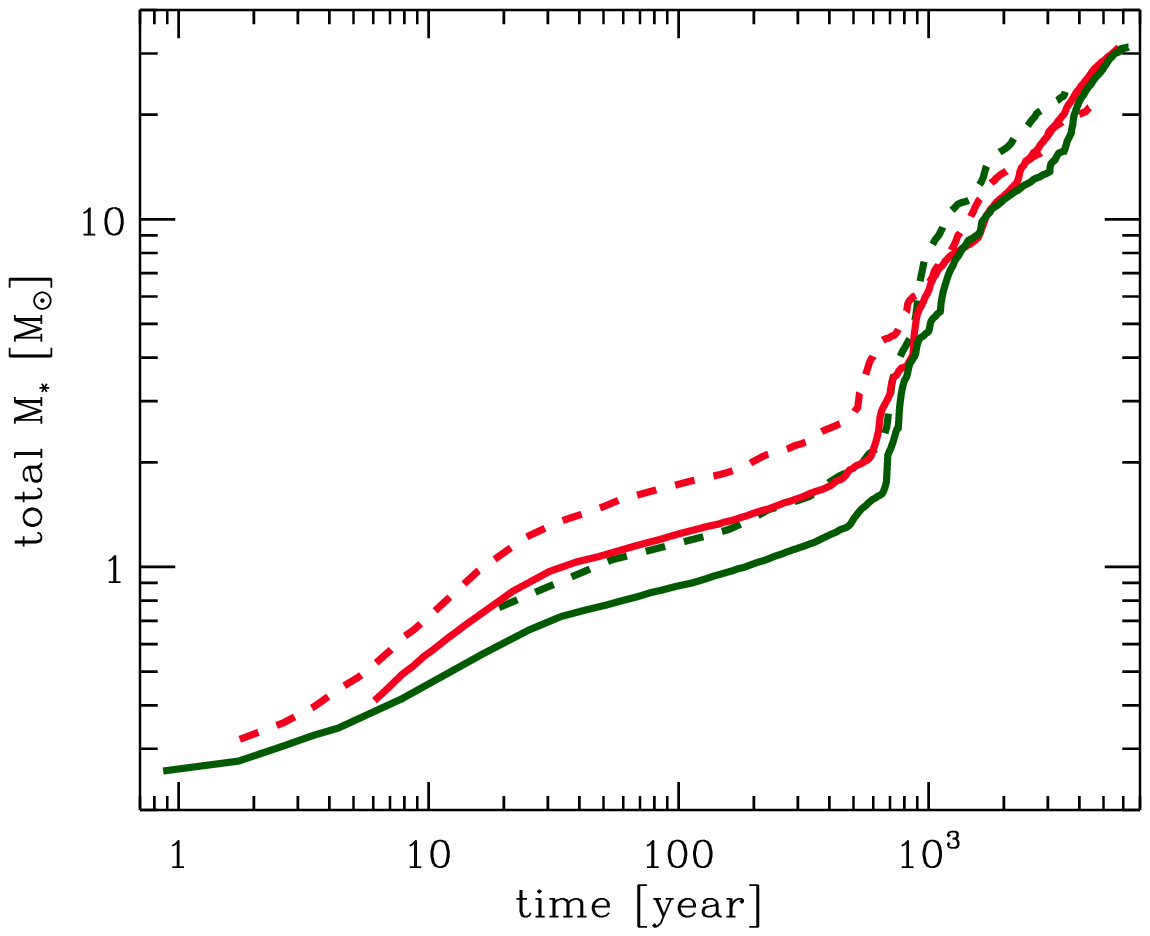}}
\put(5.5,0.0){\includegraphics[width=5.7cm]{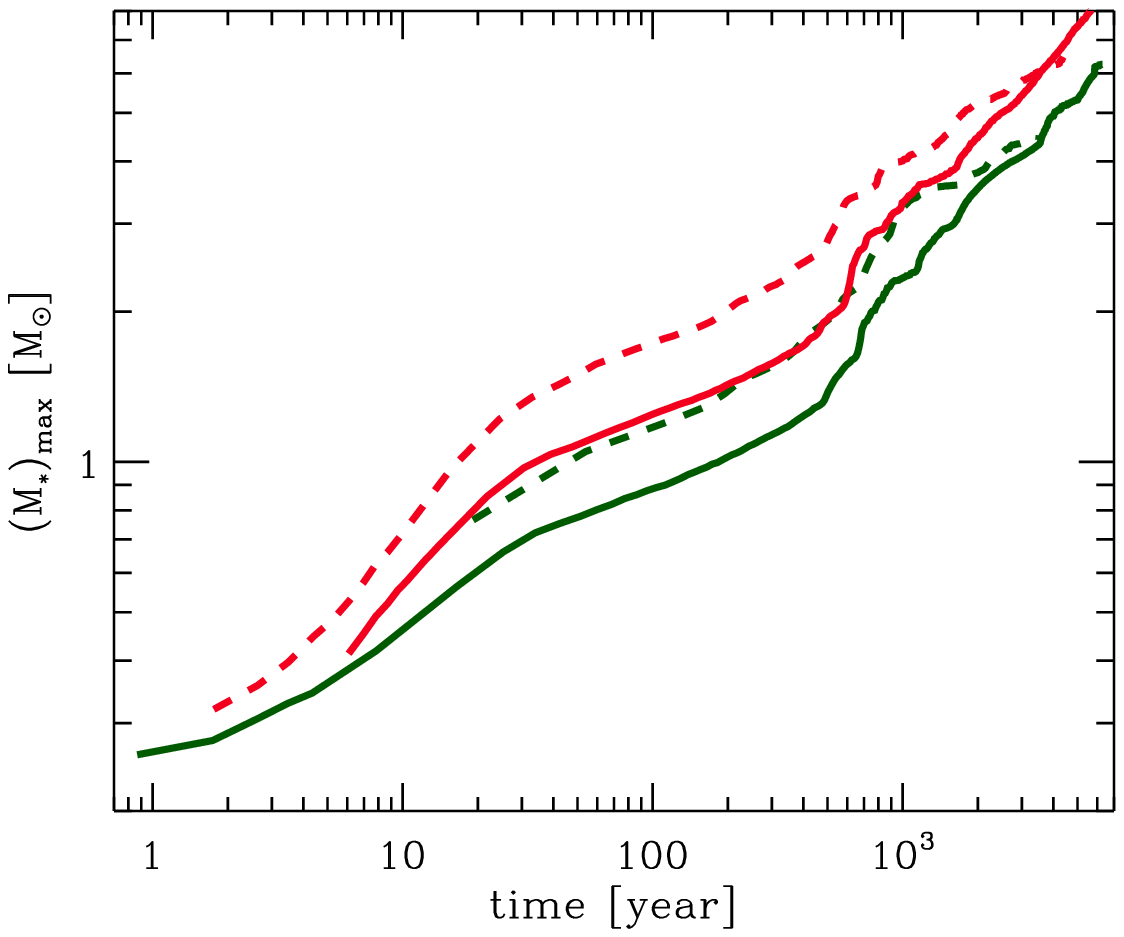}}
\put(4.5,6.1){\large A}
\put(10.5,6.1){\large B}
\put(4.5,1.1){\large C}
\put(10.5,1.1){\large D}
\put(16.5,1.1){\large E}
\end{picture}

\caption{\label{imf} 
Time evolution of the protostellar system: Accretion rate as the function of time (A) and total mass in sink particles (B). The total mass of all the sinks particles (C) and the most massive sink particle (D) are plotted as function of time. The simulations employing FH07
produce more fragmentation than ABN02 (E).
}

\end{figure*}

Figure \ref{imf}A shows the time evolution of total mass accretion rate onto the sinks (same as Figure \ref{vel}). The plot shows that mass accretion rate for ABN02 is larger than the FH07 rate. For both rates, dM$_*$/dt decreases with time. After $\approx 500$ years, further sink particles form, and the total accretion rate increases again, however, now with large temporal variations. In Figure \ref{imf}B, we plot the mass accretion rate as a function of total sink mass. 
Figure \ref{imf}C shows the time evolution of the total mass in sinks and Figure \ref{imf}D 
shows the maximum sink mass in the period over which the sink formation occurs. 

We see that the simulations employing the ABN02 rate produce a maximum sink mass that is  consistently above the FH07 rate, for a given halo. However, the difference is less than a factor of 2 and is similar to the inter-halo difference. When we examine Figure \ref{imf}E, which shows the total number of sink particle versus the total mass in all sinks, we see a similar behaviour: the differences between the runs with different rates can be a factor of two. Although in this case the halo-to-halo differences are less pronounced.

We caution the reader that our simulations do not include feedback effects from nascent protostars. We therefore restrain ourselves from following the dynamical evolution of the embedded stellar system over very long timescales. We end our calculations before we expect that radiative and mechanical feedback changes the thermal as well as chemical state -- and as a consequence -- also modifies the  fragmentation behavior  of the remaining gas  (see, e.g., the discussion in \citealt{wa08}, \citealt{whm10}, \citealt{hoyy11},  or \citealt{sgb12}).

\section{Summary}
We have investigated the effects of different proposed 3-body H$_2$ formation rates during the collapse of primordial star-forming clouds and analysed their influence on the resulting fragmentation behaviour of the gas. We compared the rates proposed by \citet{abn02} and by  \citet{fh07}, which span two orders of magnitude. We follow the chemical and thermal evolution of the primordial gas in two different minihalos in order to assess differences introduced by varying halo parameters. We find that the uncertainty in the 3-body H$_2$ formation rates leads to the differences in the chemical state of the gas which is more important for the thermal evolution of the clouds than any differences in the halo properties. The halo to halo variation 
affects the dynamical evolution of the gas and mass-accretion more than the 
chemical uncertainties. 

We also find that the simulations employing the \citet{abn02} rate to produce on average fewer and more massive fragments compared to those calculations using the \citet{fh07} rate (consistent with the
results of \citet{bsg14} which however did not use sink particles). This difference of roughly a factor of two, however, is comparable to the halo-to-halo variation. 
This is similar to the results reported by  \citet{tcggakb11} with   added advantage of using sink particles.

In summary, we find that the variations introduced to the mass distribution of primordial stars introduced by the uncertainties in the three-body H$_2$ formation rate is of similar order to the fluctuations introduced by difference in the halo properties.

%
%
%
%
%

\bigskip
The authors 
wish to thank Thomas Greif for providing the data of the minihalos 
on which our analysis is based. The authors also acknowledge Simon Glover, 
Andrea Ferrara, Kazukai Omukai, Mahavir Sharma for helpful comments. J.D. is grateful 
to the Raman Research Institute for financial support and hospitality. 
All the 
computations described here were performed on the HPC-GPU Cluster Kolob (funded 
by Heidelberg University and Deutsche Forschungsgemeinschaft). 
RSK acknowledges support from the European Research Council under the European Community's Seventh Framework Programme (FP7/2007-2013) via the ERC Advanced Grant `STARLIGHT: Formation of the First Stars' (project number 339177).

\footnotesize{

}
\end{document}